\documentclass[final,3p,times]{elsarticle}

\usepackage{graphicx}
\usepackage{epstopdf}
\usepackage[amssymb]{SIunits}
\usepackage{color}

\usepackage{amssymb}

\journal{Compte rendues}

\begin{document}

\begin{frontmatter}



\title{Aggregation of Red Blood Cells: From Rouleaux to Clot Formation}


\author{C. Wagner $^{a}$, P. Steffen $^{a}$, S. Svetina $^{b}$  }

\address{$^{a}$ Experimentalphysik, Universit\"at des Saarlandes, Postfach
151150, 66041 Saarbr\"ucken, Germany

$^{b}$ Institute of Biophysics, Faculty of Medicine, University of Ljubljana, and Jo\v{z}ef Stefan Institute, Ljubljana, Slovenia}

\begin{abstract}
Red blood cells are known to form aggregates in the form of rouleaux. This aggregation process is believed to be reversible, but there is still no full understanding on the binding mechanism. There are at least two competing models, based either on bridging or on depletion. We review recent experimental results on the single cell level and theoretical analyses of the depletion model and of the influence of the cell shape on the binding strength. Another important aggregation mechanism is caused by activation of platelets. This leads to clot formation which is life saving in the case of wound healing but also a major cause of death in the case of a thrombus induced stroke. We review historical and recent results on the participation of red blood cells in clot formation.

\textbf{R\'esum\'e}

Il est bien connu que les globules rouges forment des agr\'{e}gats, connus sous le nom de « rouleaux ». Il est souvent admis que ce ph\'{e}nom\`{e}ne d'agr\'{e}gation est r\'{e}versible, mais l'\'{e}lucidation pr\^{e}cise des m\'{e}canismes \`{a} l'{\oe}uvre dans le processus conduisant \`{a} la liaison entre globules rouges est loin d'\^{e}tre achev\'{e}. Il existe dans la litt\'{e}rature au moins deux mod\`{e}les distincts, l'un est bas\'{e} sur la formation de ponts mol\'{e}culaires et l'autre sur la notion de d\'{e}plétion. Nous passons en revue les r\'{e}sultats exp\'{e}rimentaux r\'{e}cents à l'\'{e}chelle cellulaire et analysons le modèle th\'{e}orique bas\'{e} sur la notion de d\'{e}plétion. Nous discuterons l'influence de la forme cellulaire sur la force de liaison. Un autre m\'{e}canisme d'agr\'{e}gation jouant un r\^{o}le important in vivo est celui associ\'{e} à l'activation plaquettaire. Ceci peut conduire à la formation de caillots sanguins, processus vital lorsqu'il s'agit de cicatrisation de blessures, mais qui peut \^{e}tre \'{e}galement fatal, constituant une cause majeure de d\'{e}cès, lorsqu'il s'agit de thrombose.

\end{abstract}


\end{frontmatter}



\section{Introduction}
\label{Introduction}
There are at least three major classes of aggregation of red blood cells (RBCs). First the so-called rouleaux formation that is caused by the plasma macromolecules and that is supposed to be a reversible aggregation process, second there are several indications that RBCs might become actively adhesive in the presence of a blood clot. And finally there are many pathological cases like malaria or sickle cell diseases where red blood cells are known to form large aggregates that hinder the flow of blood \cite{Hebbel1980, Shelby2003, Barabino2010}. This contribution will cover the first two of these three cases. In a recent book \cite {Baskurt2012} Baskurt et al. gave a comprehensive overview on the first topic. Here we intend to express some fresh views about the interplay between the effects on the aggregation and coagulation of the solution and of the intrinsic properties of RBCs. Most of the recent findings are largely based on the latest progresses in the application and analyses of single cell measurements.

\begin{figure} [ht!]
\centering
\includegraphics[width=0.5\linewidth]{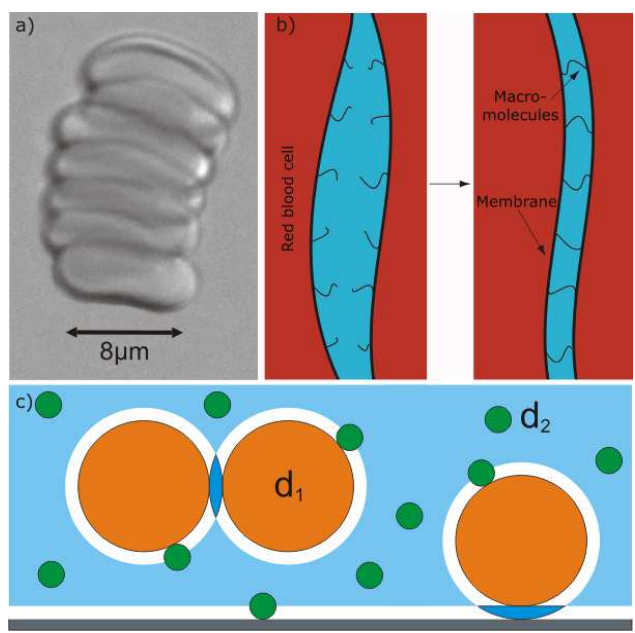}
\caption{a) From ref. \cite{Steffen2013}. Snapshot of a rouleaux of 7 RBCs in a dextran solution. b) Schematic illustration of how macromolecular bridging leads to intercellular adhesion. Macromolecules adsorb onto the RBC-membrane and are able to bridge the adjacent cell. c) Illustration of the depletion phenomenon in a binary colloidal system. The depletion layer of the colloids with diameter $d_1$ (orange) are marked white. When two depletion layers overlap, a volume $\delta V>0$ (dark blue) is released, which is additionally available for the smaller macromolecules (green) with diameter $d_2$.}
\label{fig:scetch_bridging}
\end{figure}

Human RBCs in blood samples of healthy donors have a tendency to form aggregates that look similar to a stack of coins \cite{Baskurt2012,Chien1973,Chien1987,Fahraeus1929}. These linear structures are called rouleaux. Figure \ref{fig:scetch_bridging}a shows a single rouleau in autologous plasma consisting of seven individual RBCs. The number of RBCs per rouleau can vary and branching into two rouleaux can occur. The picture of the aggregate shown in Fig. \ref{fig:scetch_bridging}a was taken under static conditions meaning that no flow was applied to the sample. The attractive forces involved in the creation process of rouleaux are relatively weak. Hence, it is possible to dissolve rouleaux into smaller fractions or even into single cells by applying a sufficient shear force \cite{Rampling1990,Schmid-Schönbein1969}. This leads to the pronounced shear thinning of blood. At low shear rates large aggregates lead to an increased viscosity, but by increasing the shear rate the aggregates break up and the viscosity decreases to a constant high shear rate value. For RBCs without  macromolecules and thus no rouleaux formation shear thinning is much less pronounced and results from deformation and orientation of the RBCs only \cite{Danker2007}. There exist two competing theories to explain the aggregation mechanism. They are based either on a model of bridging or on depletion (Fig. \ref{fig:scetch_bridging}b and c).

Formation of rouleaux is also affected by the intrinsic properties of RBCs such as the elastic behavior of the RBC membrane which contributes to the resistance of RBCs to aggregate. It is therefore of crucial importance to understand the interplay between membrane elastic energy and shapes of the agglomerate.

In general, RBCs can pass capillaries with diameters smaller than their cell size, but if clotting occurs, the flow might come to a complete stop. In the case of wound healing, clotting is life-saving, but in a healthy vessel a thrombus might lead to a stroke, the main cause of death in the developed world. The blood coagulation process is a complex process that involves many collective players and factors. In a brief manner, the coagulation process starts with the activation of blood platelets (e.g. at a damaged vessel wall). The activated platelets release a variety of messengers and growth factors in order to recruit other cells to participate in blood clot formation. Obviously, RBCs are a major part of the thrombus, but it is commonly believed that they are simply trapped in the fibrin network due to their prevalence in the blood \cite{Carr1990}. Therefore, the role of RBCs is always assumed to be completely passive. As early as a century ago, the first clinical studies described a correlation between a decreased concentration of RBCs, i.e. haematocrit value, and longer bleeding times \cite{Duke1910}. These early results have been confirmed by a number of clinical investigations \cite{Andrews1999,Horne2006}.  Considering the high percentage of RBCs in the blood and the clinical indications, it is evident that a profound understanding of the role of RBCs in clotting is crucial. An adhesion of RBCs to platelets -- and not to each other -- was assumed to be of principal importance \cite{Horne2006}. A co-adhesion of healthy RBCs was not assumed until recently Kaestner et al. \cite{Kaestner2004} suggested a signaling cascade that predicts an active participation of RBCs in blood clot formation. In first studies \cite{Nguyen2010,Steffen2011} the active co-adhesion of RBCs could be shown, but statistically significant and quantitative data are rare.

\section{Rouleaux formation}
\label{Rouleaux formation}
Rouleaux are caused by the presence of macromolecules, such as fibrinogen in blood plasma. In both, coagulation and aggregation the fibrinogen plays a crucial role \cite{Baskurt2012}. In coagulation it is converted into fibrin that is polymerized to form a mesh in which platelets and RBCs are trapped \cite{Brummel1999}. Similar rouleaux formation can also be induced by re-suspending the RBCs in physiological solutions containing neutral macromolecules such as dextran \cite{Pribush2007}. However, without any macromolecules, e.g. RBCs in a simple salt solution, no aggregation occurs. The fibrinogen mediated aggregation of RBCs increases consistently with increasing fibrinogen concentration \cite{Marton2001}, whereas the dextran-mediated aggregation of RBCs reaches a maximum at a certain dextran concentration. The strength of the aggregation depends not only on the dextran concentration, but also on the molecular weight of the dextran (i.e., the radius of gyration of the dextran) \cite{Chien1987,Brooks1988}.
To this day the mechanisms involved in RBC aggregation have not been fully understood. There are two coexisting models that try to explain the rouleaux formation of RBCs: the bridging model and the depletion model. In the bridging model, it is assumed that fibrinogen or dextran molecules non-specifically adsorb onto the cell membrane and form a "bridge" to the adjacent cell \cite{Brooks1988}. In contrast, the depletion model proposes the opposite. In this model, aggregation occurs because the concentration of macromolecules near a RBC surfaces in close proximity is depleted compared to the concentration of the bulk phase, resulting in a net "depletion" force.

The bridging model assumes that large macromolecules adsorb onto the cell surface and thereby bridge two adjacent cells. When these bridging forces exceed the disaggregation forces such as electrostatic repulsion, membrane strain and mechanical shearing, aggregation occurs \cite{Chien1987,Brooks1988,Brooks1973,Chien1973b,Snabre1985}. More precisely, studies that focused on the inter cellular distance \cite{Chien1973} of two adjacent cells showed that the intercellular distance is less than the size of the hydrated molecules. This leads to the assumption that the terminal portions of the flexible polymers are adsorbed onto the surfaces of adjacent cells, resulting in a cell-cell adhesion (see Fig. \ref{fig:scetch_bridging}). Thereby, the cell-cell distance increases with increasing polymer size but is always smaller than the diameter of the hydrated polymer \cite{Chien1975}.

In the depletion model aggregation occurs because the concentration of macromolecules near a RBC surface in close proximity is depleted compared to the concentration of the bulk phase, resulting in a net "`depletion"' force. A first explanation of depletion forces was given by Asakura and Oosawa \cite{Asakura1958}, who discovered that the presence of small spheres (i.e., macromolecules) can induce effective forces between two larger particles if the distance between them is small enough. The origin of these forces is purely entropic. When two large plates are immersed in a solution of rigid spherical macromolecules and the distance between the inner surfaces of these two plates is smaller than the diameter of solute macromolecules, none of these macromolecules can enter the space between the plates and this space becomes a phase of the pure solvent. Therefore, a force equivalent to the osmotic pressure of the solution of macromolecules acts on the outer surfaces of these plates. Such a force also appears between two spherical particles if the distance between the two large particles decreases to less than the size of the surrounding macromolecules (Fig. \ref{fig:scetch_bridging}c). In such a system a so-called depletion layer surrounds the large particles where the centers of the smaller particles cannot enter. Consistently, in that depletion layer the concentration of macromolecules becomes depleted compared to that of the bulk. The thickness of this depletion layer equals the radius of the smaller particles. When overlapping between two depletion layers occurs, an additional free volume is available for the smaller particles causing an increase in entropy and hence a decrease in Helmholtz's free energy leading to an effective osmotic pressure causing an attractive force between the large particles. In diluted solutions below the overlap concentration \cite{DeGennes1979} polymers can be treated as rigid spheres with a radius $d_2/2$ matching the radius of gyration of the polymer \cite{Ohshima1978}.

\subsection{Bridging versus depletion}
Over the past, there have been studies in support of both theories. Studies in favor of the bridging model dealt with either aggregation induced by nonspecific binding of macromolecules \cite{Brooks1980,Chien1975} or by specific binding mechanisms \cite{Lominadze2002}. The determination of macromolecular adsorption of polymers and proteins to RBCs are subject to a lot of possible artifacts and consequently the interpretation of existing data is difficult \cite{Janzen1989,Janzen1991}. Despite a lot of efforts to quantify macromolecular binding, conclusive data is still lacking.
On the other side, several studies favoring the depletion model have been published \cite{Armstrong2004,Baeumler1996,Neu2002}. Neu and Meisselman \cite{Neu2002} adopted the depletion concept and applied it on the aggregation of red blood cells and developed a theoretical description of the acting forces. They also predicted quantitative values of the interaction energies.

\subsection{Theoretical description of depletion based aggregation of RBCs}
\begin{figure} [ht!]
\centering
\includegraphics[width=1.1\linewidth]{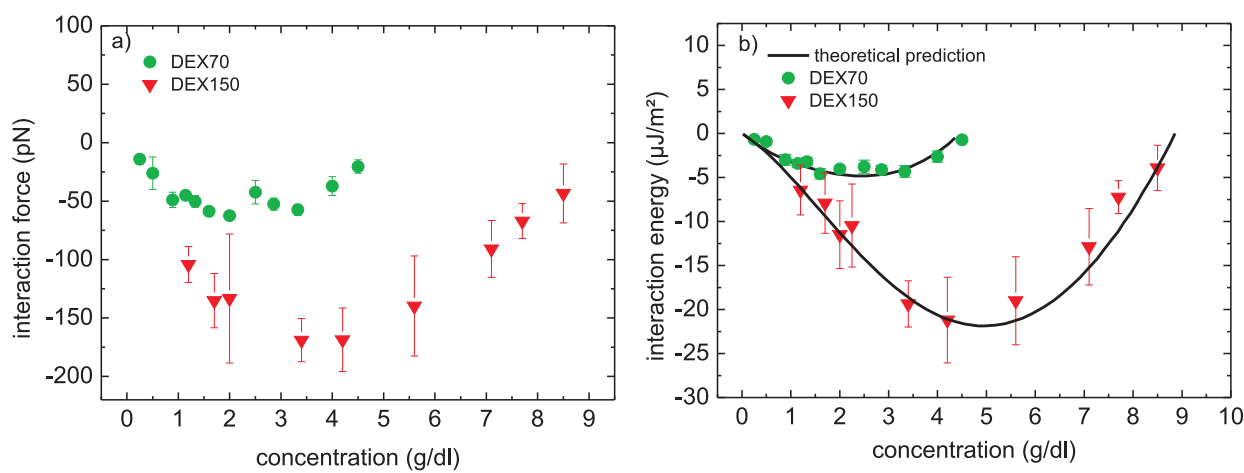}
\caption{From ref. \cite{Steffen2013}. a) The measured adhesion forces for different concentrations of dextran (circles:  dextran 70, triangles: dextran 150). The maximum interaction strengths were observed at $2 \gram/\deci\litre$ (dextran 70) and $4 \gram/\deci\litre$ (dextran 150). b) Dependence of the interaction energy of two red blood cells for different concentrations of dextran. The solid line represents the curve calculated by Neu and Meisselman \cite{Neu2002}.}
\label{fig:interaction_curve}
\end{figure}

The nature of the interaction forces depends on the surfaces of the adhering objects. The surface of the RBCs is strongly influenced by the soft layer attached to the membrane, called glycocalyx. Therefore, one has to take the possibility into account that the RBCs also interact via steric interaction due to the overlapping glycocalyces. Due to the high electrostatic repulsion, cell-cell distances at which minimal interaction energies occur (i.e. maximal adhesion strength) are always greater than twice the thickness of the glycocalyx \cite{Neu2002}. Thus, steric interactions can be neglected for the case of RBCs and only depletion and electrostatic repulsion have to be considered. Neu and Meisselman \cite{Neu2002} computed the effects of bulk phase polymer concentration on interaction energy. They identified the softness and the consecutive penetration of dextran molecules into the soft RBC surface as crucial when it comes to the development of the characteristic bell-shaped relations of the interaction energy in dependence on the dextran concentration (Fig. \ref{fig:interaction_curve}) while without surface penetration one finds a linear dependence of the interaction energy on the dextran concentration. At first sight these bell-shaped curves might seem counterintuitive because considering depletion interaction one expects a consecutive increase in the interaction energy with increasing polymer concentration. The reason for these bell-shape nature of the curves lies in the increasing penetration depth of the polymers into the glycocalyx with increasing bulk polymer concentration and in the decreasing depletion layer thickness with increasing bulk polymer concentration.
In order to compare the computed model with experimental data, Neu and Meisselman \cite{Neu2002} took data from Buxbaum et al. \cite{Buxbaum1982}. There, a micro pipette based approach was chosen to measure the surface affinities of a RBC membrane vesicle with an intact RBC. Neu and Meisselman \cite{Neu2002} varied the penetration constant until the calculated peak interaction energy for dextran $70\,\kilo\dalton$ or dextran $150\,\kilo\dalton$ equaled the value reported by Buxbaum et al. \cite{Buxbaum1982}.

\subsection{Single cell force measurements}

The technique of single cell force spectroscopy (SCFS) was used to measure the interaction energies between human red blood cells as functions of the molecular weight and concentration of dextran \cite{Steffen2013}. The dextrans used were dextran70 (DEX70 with a molecular weight of $70 \,\kilo\dalton$), and dextran150 (DEX150 with a molecular weight of $150\, \kilo\dalton$) from Sigma-Aldrich. The measurements were conducted at the single cell level and were compared to the predicted values of Neu and Meisselman \cite{Neu2002}. An atomic force microscope (AFM) (Nanowizard 2, equipped with the CellHesion Module with an increased pulling range of up to $100\, \micro\meter$, JPK Instruments, Germany) was used to conduct single cell force spectroscopy measurements \cite{Friedrichs2010}. In the course of the experiment, a single RBC was attached to an AFM cantilever by appropriate functionalization. Cell Tak$^{TM}$ (BD Science) was used to bind a cell to the cantilever. After cell capture, the cantilever was lowered onto another cell, and the adhesion force and adhesion energy were measured. The retraction curve is typically characterized by the maximum force required to separate the cells from each other and adhesion energies are calculated by computing the area under the retraction curve of the force distance curve. The interaction energies (more precisely, the interaction energy densities of two RBCs) are calculated by dividing the measured adhesion energies by the contact areas of the adhering cells using a value of $50.24\, \micro\meter^2$ derived from the maximum radius of RBCs. Figure \ref {fig:interaction_curve}a and b show the dependence of the adhesion force and the interaction energy on the concentration of the dextran used.

Since little statements regarding adhesion forces are made by the literature, it is important to consider the interaction energies, too. The measured values of the adhesion energy can now be compared to the values predicted by the theory \cite{Neu2002}. The excellent agreement of the SCFS measurements with the theoretical description give more rise to the assumption that the driving force in rouleaux formation is rather depletion induced than bridging induced. As could be seen in contact time dependence measurements, the adhesion energies increase with increasing contact time. This could be due to bridging, but there is not enough data to conclusively decide that. With the present data it appears that the rouleaux formation, at least at the beginning, is purely depletion mediated.

\subsection{The role of RBC shape transformations in the RBC aggregation processes}
Aggregated RBCs exert spatial constraints on each other and therefore their shapes differ from the shapes that they attain when free. As the shape of the unconstrained, free cell corresponds to the smallest possible membrane elastic energy, in an aggregated RBC this energy is larger. RBC aggregation can thus only occur if the total energy decrease due to the attraction between cells exceeds the corresponding increase of the elastic energy of their membranes. The knowledge about the factors that affect RBC membrane elastic energy is therefore for the understanding of the RBC aggregation process equally important as the knowledge about the factors involved in cell-cell interactions. In this subsection we shall discuss the relationship between RBC shapes in RBC aggregation process at the macroscopic level. In order to reveal the essential features of the role of RBC shapes in the aggregation process, we shall give an account mainly on the works on the simplest possible (minimal) model of the RBC aggregation. Within this model the RBC is assumed to behave analogously to the behavior of lipid vesicles with homogeneous bilayer membranes, therefore we shall review also some works on the adhesion of vesicles. When applicable we shall also comment on the limitations of the treated minimal model.

The general feature of the adhered RBC is that its membrane is divided into zones which are in contact with surfaces of other cells and zones which are in contact only with the surrounding solution. The adhesion energy pertains only to the zones that are in contact. In the minimal-prototype model of the RBC aggregation process it is assumed that membranes involved are laterally homogeneous and that it is possible to define an adhesion energy ($W_a$) which is the product of the contact area ($A_c$) and an adhesion constant ($\Gamma$), e.g. measured in $J/\mu m^2$ (see Fig. \ref{fig:interaction_curve}):
\begin{equation}
W_a = - \Gamma A_c
\end{equation}
Shapes of adhered or aggregated RBCs correspond to the minimum of the energy functional which in addition to Eq. 1 involves the elastic energy of its membrane. The latter is the sum of the elastic energy of the bilayer part of the RBC membrane and the elastic energy of its membrane skeleton. The minimal model of the RBC aggregation takes into consideration only the local \cite{Helfrich1973} and non-local \cite{Helfrich1974,Evans1980b,Svetina1985,Miao1994} bending energy terms of the RBC membrane bilayer, expressed, respectively, as
\begin{equation}
W_b= \frac{1}{2}k_c\oint(C_1+C_2-C_0)^2dA+\frac{1}{2}\frac{k_r}{h^2A_0}(\Delta A-\Delta A_0)^2
\end{equation}
where $k_c$ is the bending modulus of the bilayer, $C_1$ and $C_2$ are the principal curvatures of the membrane, and $C_0$ is the spontaneous (preferred) curvature of the membrane. $k_r$ is the nonlocal bending modulus, $\Delta A$ is the difference between the areas of the outer and the inner leaflets equal to $h\oint(C_1+C_2)dA$ where $h$ is the distance between neutral surfaces of the membrane leaflets, $\Delta A_0$ is the equilibrium (preferred) difference between the areas of the outer and the inner leaflets which essentially depends on the difference in the molecular occupancy of the two layers, and $A_0$ is the equilibrium (preferred) membrane area. The value of the local bending constant for the RBC membrane is $k_c = 2 x 10^{-19} J$ while $k_r$ is about twice that large \cite{Hwang1997}. If the lateral tensions of the two bilayer leaflets equilibrate, e.g. by way of the flip-flop transport of lipid molecules, the non-local bending term vanishes.
The general problem to be solved in the RBC aggregation phenomenon is to find the minimum of the sum of Eqs. 1 and 2 for all cells that form a given aggregate. Technically the problem is a generalization of the shape determination of a single unconstrained vesicle/cell which corresponds to the minimization of Eq. 2. The latter topic has been in the past reviewed comprehensively \cite{Seifert1997,Sventina2009}. Shapes can be theoretically determined by solving the corresponding shape equation which gives beside the coordinates of all membrane points also the corresponding principal curvatures. For the present discussion it is of interest to note that at the reduced volume (volume divided by the volume of a sphere with the same membrane area) of the RBC which is about 0.6, the shape that corresponds to the minimum of the local bending energy is a discocyte and thus coincides with the shape of the RBC in its resting state. The decrease of either spontaneous curvature $C_0$, equilibrium area difference $\Delta A_0$, or both causes RBC to transform into an oblate stomatocyte. The increase of these quantities tends to make the shapes prolate, however, RBCs instead transform into spiculated echinocytes which has been interpreted by taking into account the elasticity of the RBC membrane skeleton \cite{Lim2002}. This means that the minimal model of the RBC aggregation is applicable only when RBC shape transformations do not involve too strong changes of membrane principal curvatures.

When measuring the relative effects of the adhesion energy it is appropriate to look into how its magnitude compares to the membrane bending energy. A suitable parameter that was chosen to quantify this comparison is the ratio between the adhesion energy corresponding to the area of the cell membrane $\Gamma 4\pi R_0^2$ (where $R_0$ is the radius of the sphere with the RBC membrane area $A_0$) and the bending energy of a spherical membrane ($8\pi k_c$),
\begin{equation}
\gamma = \Gamma R_0^2/2k_c.
\end{equation}
For $\gamma \ll 1$ the adhesion is negligible and for $\gamma \gg 1$ it prevails. Eq. 3 also indicates that at given values of the adhesion constant and the reduced volume the effects of the adhesion are more pronounced in larger cells.

In treating the RBC shape behavior in the aggregation processes it is necessary to take into account that we are dealing with the adhesion that occurs between two flexible surfaces. This is important to note because most works on vesicle and cell adhesion deal with their adhesion to rigid surfaces. When RBC is adhered to a rigid surface its contact zone attains the shape of the surface. It is then only necessary to determine the shape of the non-adhering membrane zones which can be done by solving the shape equation for those membrane sections. The variables in the corresponding shape determination are the membrane principal curvatures and the area of the adhesion zone. For solving the shape equation it is necessary to know the values of principal curvatures at the zone boundaries. For the vesicle adhesion to the flat rigid surface this condition was derived \cite{Seifert1990} and reads for axisymmetrical shapes
\begin{equation}
	c_m = (4\gamma)^{1/2},	
\label{cm}
\end{equation}
where $c_m$ is the reduced value of the principal curvature along the meridians, $c_m = R_0C_m$. When the adhesion occurs between two flexible surfaces which is the case in the RBC-RBC adhesion, the shape equation has to be solved for all membrane zones. For two axisymmetric adhered cells with membranes of equal bending constants the generalized boundary conditions read \cite{Derganc2003}
\begin{equation}
	\Delta c_m = (2\gamma)^{1/2},	
\label{deltacm}
\end{equation}
where $\Delta c_m$ is the difference between the reduced principal curvature along meridians of the free and adhered parts of the membrane. It was shown that also for vesicles of a general symmetry the squared curvature jump $4\gamma$ demanded by the rigid surface version (Eq. \ref{cm}) is shared in equal parts ($2\gamma$) between the two membranes \cite{Deserno2007}. The problem of boundary conditions for adhered soft membranes was recently also comprehensively treated by Agrawal \cite{Agrawal2011}.

The classical example of RBC aggregation is the formation of the rouleau. Rouleaux were first treated by the minimal model by Skalak et al. \cite{Skalak1981}. They considered the adhesion zone to be flat. Consequently, each cell in such a linear array was symmetrical with respect to the mirror equatorial plane. More recently the question was raised if the cells in the rouleau can have no such symmetry \cite{Derganc2003}. Rouleau was treated as an infinite chain of identical axisymmetrical cells so that it was possible to apply periodic boundary conditions and determine the RBC shape by treating a single cell. It was shown that there is the transition from mirror symmetrical to asymmetrical shapes that occurred by lowering the reduced volume or/and by increasing the adhesion constant. The shape parameter that favors the asymmetrical shapes was shown to be also the preferred difference between the areas of the bilayer layers ($\Delta A_0$). The question about possible symmetry of adhered RBCs was reexamined by applying numerical methods for RBC shape determination and thus removing the restriction about the axial symmetry \cite{Ziherl2007}. RBC doublet was studied and it was shown that the RBCs in a doublet indeed exhibit a transition from axially symmetrical to non-axisymmetrical shapes. When the reduced adhesion strength $\gamma$ is small but larger than the threshold for adhesion, the contact zone is planar and circular as if each of the vesicles would stick to the rigid surface, and the shapes of the two vesicles are the same. If the adhesion strength is increased further, the stable doublet consists of identical vesicles joined in a sigmoidal, S-shaped contact zone with an invagination and a complementary evagination on each vesicle. There are two features of these doublets which may have a role in the formation and the outlook of rouleaux. At some intermediate adhesion strengths the membranes of the two constituent cells of a doublet are in the contact zones rather curved, whereas their outer surfaces are practically flat. Such doublets could adhere to each other without an additional increase of the membrane elastic energy. Because the cells in a doublet with a sigmoidal contact zone are shifted away from the rouleau long axis in opposite directions, in the so formed rouleaux the cells arrange in a zig-zag manner. Such cell arrangements have been indeed observed \cite{Skalak1981,Kirschkamp2008}.

The study of doublet shapes on the basis of the described minimal model of RBC aggregation has been employed to envisage different adhesion regimes with respect to the reduced adhesion strength $\gamma$ and characterized by a specific type of aggregate \cite{Svetina2008}. In the regime right above the adhesion threshold ($0.3 \lesssim  \gamma \lesssim  ~2$) the predominant aggregate shape is the flat-contact doublet. In the weak adhesion regime ($~2 \lesssim  \gamma \lesssim  ~4$) doublets aggregate into the zig-zag rouleaux while in the strong adhesion regime ($~6 \lesssim  \gamma \lesssim  ~10$) predominate sigmoid-contact round doublets. At still higher values of the adhesion parameter ($\gamma \gtrsim ~15$) the cells would favor to aggregate into rounded clumps. In general it looks that rouleaux are formed in the regime in which there is still a reasonable balance between the adhesion and bending energies.

Because of many limitations of the minimal model for RBC aggregation the above conclusions are only qualitative. First of all, the minimal model does not consider the area expansivity and shear energy terms of RBC membrane skeleton \cite{Mukhopadhyay2002,Kuzman2004} which, as already stated, becomes important especially in shape transformations to or from the shapes exhibiting high curvatures. It can be concluded that the aggregation behavior depends on the initial RBC shapes and is different if they are for example discocytes or echinocytes. The next possible future generalization of the minimal model is the effect of adhesion on the lateral segregation of membrane components. In the aggregation where RBCs adhere to other blood cells, it will be important to derive the boundary conditions analogous to Eq. \ref{deltacm}  which would take into consideration that the two adhering membranes have different mechanical properties \cite{Agrawal2011}. There are also still unanswered problems related to the size and structure of rouleaux. RBCs are variable in their sizes and this certainly affects the sizes of rouleaux, especially it is expected that the end cells could have very specific sizes. The effects of the initial RBC shapes are of particular interest because non-discoid shapes have been found in some hereditary or other diseases. The observation of RBC shapes and their aggregation behavior may serve as possible indicators of these diseases.

\section{Platelet induced coagulation of red blood cells}


\begin{figure} [ht!]
\centering
\includegraphics[width=0.5\linewidth]{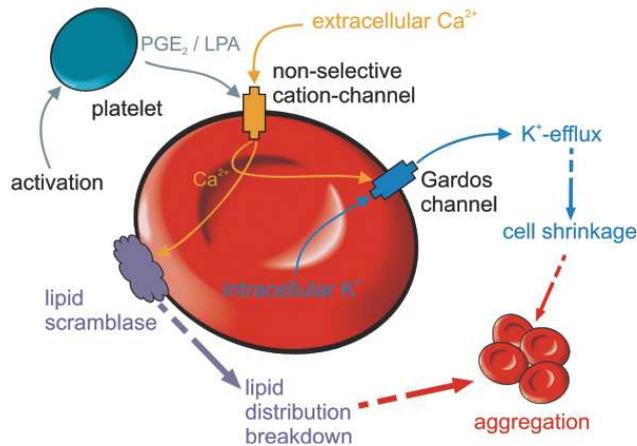}
\caption{Signaling Cascade of the LPA induced adhesion of RBCs.}
\label{fig:signalling-cascade}
\end{figure}

Coagulation is governed by a complex signaling cascade and it requires a number of essential enzymes and cofactors, so-called coagulation factors, and is finalized by the formation of the thrombus \cite{Mackman2008}. The platelets themselves are the most important cells in the coagulation process, inside those platelets the so-called scramblase protein is activated. The scramblase activation results in a degradation of the asymmetrical distribution of phospholipids in the lipid bilayer of the RBC membrane. This leads to an exposition of the negatively charged phospholipid phosphatidylserine (PS) in the outer leaflet of the RBC membrane. This PS is under suspicion to play an important role in blood coagulation since it might provide a catalytic surface for prothrombinase complexes \cite{Chung2007}. Much work has already been conducted on the hypothesis of an active participation of RBCs in thrombus formation, but a direct investigation of the involved adhesion forces in order to check this hypothesis was missing until recently. Optical tweezers as well as single cell force spectroscopy were used to check this hypothesis and to quantify the occurring adhesion in terms of adhesion strength \cite{Steffen2011}.

The activation of RBCs by activated platelets involves a specific signaling cascade (Fig. \ref{fig:signalling-cascade}). Kaestner and Bernhardt hypothesized a $Ca^{2+}$ influx via a non-selective cation channel (NSC) \cite{Christopherson1991,Kaestner1999,Kaestner2004} which is permeable to different mono and bivalent cations such as $Na^+$, $K^+$ or also $Ca^{2+}$. This channel is opened by physiological concentrations of prostaglandin $E_2$ ($PGE_2$) and lysophosphatidic acid ($LPA$) \cite{Kaestner2002,Kaestner2004,Kaestner2006}. These substances are released by activated platelets. The increased intracellular $Ca^{2+}$-concentration acts as a trigger mainly for two processes. First, another calcium-dependent channel (the Gardos channel) is opened by the increased calcium level \cite{Gardos1958}. Through this channel, intracellular potassium effluxes out of the cell, followed by a shrinkage of the cell \cite{Li1996,Lang2003}. The second, more important, process is that the lipid scramblase protein is activated which has a profound consequence: the breakdown of the asymmetrical lipid distribution between both leaflets \cite{Basse1996,Williamson1992,Williamson1995,Woon1999,Dekkers2002}. In this way, the negatively charged phospholipid phosphatidylserin (PS), usually exclusively present in the inner leaflet, is transported into the outer leaflet \cite{Nguyen2010}. It was shown that the PS-exposure is the key player in the adhesion of RBCs to endothelium cells \cite{Closse1999, Mandoori2000}. It is also worth to mention that besides e.g. diseases like malaria or sickle cell disease there are a various number of processes that are related to PS exposure in the outer leaflet \cite{Luvira2009,Eldor2002,Taher2008}. Examples are cell adhesion in general, cell fusion \cite{Tullius1989}, phagocytosis \cite{Zwaal1997,Woon1999,Zwaal2005,Tanaka1983,Schroit1985} and  apoptosis \cite{Messmer2000,Zwaal2005}.

\subsection{Red Blood Cell Stimulation with LPA}
\begin{figure} [ht!]
\centering
\includegraphics[width=0.6\linewidth]{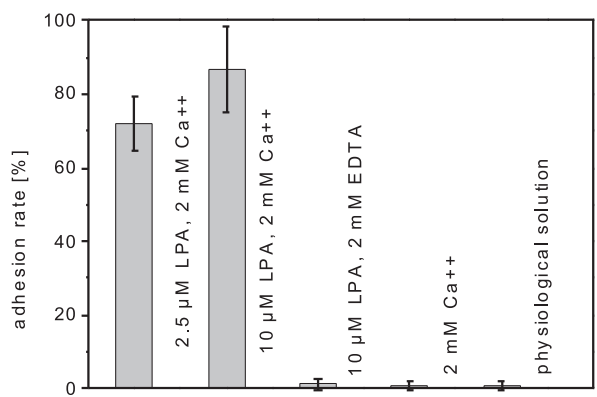}
\caption{From ref. \cite{Steffen2011}. Results of the LPA measurements conducted with optical tweezers. The gray bars represent the percentage of cells that showed adhesion. The overall number of cells tested was 60 cells per measurement. In the presence of LPA and $Ca^{2+}$, a significant number of cells showed adhesion, whereas in the control experiments, only a very small portion of the cells showed an adhesion. The results of the student's t-test, compared to the control measurement (HIS-solution), are indicated at the top of each bar.}
\label{fig:adhesion-statistic-tweezers-LPA}
\end{figure}

As pointed out in the introductory part of this section, RBCs can be stimulated by LPA, and this has been proposed to contribute to the active participation of RBCs in the later stage of thrombus formation. In order to test for altered intercellular adhesion behaviour under different conditions, cells were treated with various solutions and were investigated in microfluidics with holographic optical tweezers. Upon stimulation with $2.5\,\micro M$ LPA, the RBCs adhered to each other within 30 to 60 seconds. During the stimulation procedure, most of the RBCs remained in their discocyte shape. To exclude any dependencies on the interaction surface due to the anisotropic shape of the cells, we aimed for another condition using spherocytes \cite{Kaestner2011}. This was realized by increasing the LPA concentration to $10\,\micro M$, which is still within the physiologically observed range. The separation force could not be determined by the optical tweezers approach because it exceeds the force of the laser tweezers, which, in this particular setup, amounts to $15-25\, \pico\newton$. Consequently, at this stage, the adhesion force could only be qualified to be larger than this. Nevertheless, with this approach a statement about the general behaviour of the treated RBCs and the adhesion statistics could be reached. Cells under five different conditions were tested: (i) in a HEPES buffered solution of physiological ionic strength (PIS-solution) containing the following (in mM): 145 NaCl,7.5 KCl, 10 glucose and 10 HEPES, pH 7.4, at room temperature, (ii) PIS-solution containing $2\,\milli M$ $Ca^{2+}$ and $10\,\micro M$ LPA, (iii) PIS-solution containing $2\,\milli M$ $Ca^{2+}$ and $2.5\micro M$ LPA, (iv) PIS-solution containing  $2\,\milli M$ $Ca^{2+}$ and no LPA, and finally (v) PIS-solution containing $2\,\milli M$ EDTA and $10\,\micro M$ LPA. We used at least 60 cells per condition. The results are summarized in Fig. \ref{fig:adhesion-statistic-tweezers-LPA}.
The RBC stimulation with LPA ($2.5\,\micro M$ as well as $10\,\micro M$) in the presence of extracellular $Ca^{2+}$ led to an immediate qualitative change in the adhesion behaviour: cells stuck irreversibly to each other. In case of treatment with $2.5\,\micro M$ about $72\,\%$ of the cells tested showed an irreversible adhesion. In the case of $10\,\micro M$, which still represents a physiological concentration near or inside a blood clot \cite{Eichholtz1993}, the adhesion rate went even up to more than $90\,\%$. In the control measurement in which the extracellular $Ca^{2+}$ was chelated by EDTA and in the control measurements without any RBC treatment, almost no adhesion events could be seen.

\subsection{Quantification of the Intracellular Adhesion}
\begin{figure} [ht!]
\centering
\includegraphics[width=0.8\linewidth]{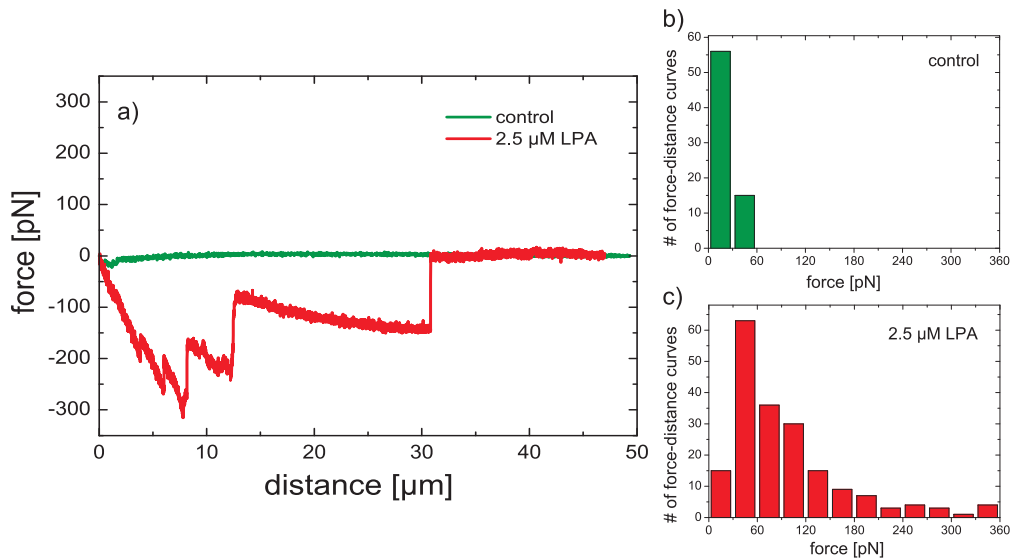}
\caption{From ref. \cite{Steffen2011}. a) Shows the combined plot of an example force-distance-curve of a control (green) and an LPA measurement (red). b) The statistics of measured adhesion forces in the control measurement (without LPA treatment) c) Recorded data of the measured forces for LPA treated RBCs.}
\label{fig:afm-LPA-control-curve}
\end{figure}

To allow a discussion of a physiological (or pathophysiological) relevance of the described adhesion process, one needs to determine the separation force. As described above, the separation force exceeds the abilities of the HOT. Therefore, single-cell force spectroscopy \cite{Friedrichs2010} was utilized to determine the force. Two different measurements were conducted: control measurements in which the cells remained untreated, and measurements in which the cells were treated with a concentration of $2.5\,\micro M$ LPA. A concentration of $2.5 \,\micro M$ was chosen for most of the measurements because it resembles most physiological concentrations. Later on, further measurements with $10\,\micro M$ LPA were conducted as well to check again for the influence of the cell shape on the measured forces.
Fig. \ref{fig:afm-LPA-control-curve}a shows two example curves of measurements with $2.5\, \micro M$ LPA treated cells (red) and untreated cells (green). In most of the measurements, a significantly changed adhesion behaviour was observable after the treatment with LPA.
Whereas the measured forces in the case of untreated cells barely exceeded $20\,\pico\newton$, in the case of LPA treatment this measured force went up to even more than $300\, \pico\newton$ in some cases. However, the measured forces strongly varied from cell to cell. In total, more than 50 cells were tested for each case and the results are summarized in Fig. \ref{fig:afm-LPA-control-curve}. The mean value of the maximum unbinding force of untreated RBCs (control, green) amounted to $28.8\, \pm 8.9\, \pico\newton$ (s.d.) (n=71), whereas in the LPA experiments, the mean value of the maximum unbinding force amounted to a much higher value of $100\, \pm 84\, \pico\newton$ (s.d.) (n=193, from three different donors), indicating a severe difference in adhesion behavior of untreated and LPA-stimulated RBCs. The occurrence of the small adhesion forces in the control measurements results probably from instrumental artefacts. Upon stimulation with such LPA, RBCs adhere irreversibly to each other. The separation force of approximately $100\,\pico\newton$ (determined by single cell force spectroscopy) is in a range that is of relevance in the vasculature \cite{Snabre1987}. Due to the relatively slow response of the RBCs upon stimulation (30 to 60 s) an initiation of a blood clot based on intercellular RBC adhesion is regarded to be irrelevant under physiological conditions. Once caught in the fibrin network of a blood clot, the adhesion process observed here in vitro may support the  solidification of the clot. This notion is supported by the aforementioned experimental and clinical investigations reporting a prolongation of bleeding time in subjects with low RBC counts \cite{Hellem1961,Livio1982,Kaestner2004,Mackman2008}.

\section{Summary}
In this review different adhesion phenomenona of red blood cells (RBC) are discussed. RBC aggregation has already been known for a long time and is the main determinant of blood viscosity. Additionally, there are many indications that RBC aggregation could play a role in thrombus formation and thrombus solidification. In order to quantify the different adhesion phenomenona, accurate force measuring tools are required.

Aggregation that is induced by the presence of plasma macromolecules or in model systems by dextran polymers is called rouleaux formation. The  underlying adhesion mechanism is not fully understood yet. Two different models were developed to explain the origin of the adhesion, either based on bridging or on depletion. There are experimental data in favour of both theories, but all of them are indirect measurements and not actual cell-cell interaction measurements. Cell-cell adhesion measurement of RBCs in their natural, discocytic shape by means of modern spectroscopic methods like optical tweezers or atomic force microscopy based single cell force spectroscopy (SCFS) were performed only recently. It turned out that the measured interaction energies are in excellent agreement with the ones predicted by the depletion theory. Therefore, based on this data, it can be concluded that the rouleaux formation is rather depletion-mediated than bridging-mediated. For longer contact times of the cells additional enhanced interaction energies could suggest an influence of bridging in the later stages of adhesion, but could not be conclusively confirmed.

Among the intrinsic RBC properties that affect their aggregation, the elastic properties of their membranes play an important role. In order for the RBC aggregates to be stable, the cell-cell attraction has to be stronger than the energy needed for the accompanying RBC shape transformation. Recent corresponding theoretical work emphasized specificities in treating the adhesion between flexible surfaces. A significant theoretical prediction was about the role of RBC doublets in the formation of rouleaux.

The platelets induced adhesion of red blood cells has been only recently studied.  A combined approach of microfluidics and holographic optical tweezers were used to test the hypothesis statistically. The arising adhesion exceeded the force capabilities of optical tweezers, thus single cell force spectroscopy was used to quantitatively investigate this adhesion phenomenon. This approach quantified the intercellular adhesion to amount to $100\, \pm 84\, \pico\newton$ in strength. Concerning the question of physiological significance, the results of HOT and SCFS were combined and it could be concluded that the LPA-induced intercellular adhesion of RBCs is of importance in the later stages of blood clotting and actively contributes to blood clot solidification.\\

\appendix

\section{Acknowledgements}
\label{}

\bibliographystyle{elsarticle-num}

\end{document}